\documentclass{aa}  

\usepackage{graphicx}
\usepackage{txfonts}
\usepackage{xcolor}

\newcommand{\red}[1]{\textcolor{black}{#1}}
\usepackage{hyperref}
\begin{document}

%   \title{Near infrared spectra of TiO$_2$ clusters for hot Jupiter atmospheres.}
\title{Infrared spectra of TiO$_2$ clusters for hot Jupiter atmospheres}
\subtitle{}

\author{J.P. Sindel
      \inst{1,2,3,4}
        \and
      Ch. Helling \inst{1,5}
      \and
      D. Gobrecht\inst{4,6}
      \and
      K.L. Chubb \inst{2}
      \and
     L. Decin \inst{4}
      }

\institute{Space Research Institute, Austrian Academy of Sciences, Schmiedlstrasse 6, A-8042 Graz, Austria\\
\email{JanPhilip.Sindel@oeaw.ac.at}
\and 
Centre for Exoplanet Science, University of St Andrews, North Haugh, St Andrews, KY169SS, UK
    \and
        SUPA, School of Physics \& Astronomy, University of St Andrews, North Haugh, St Andrews, KY169SS, UK
        \and
         Institute for Astronomy, KU Leuven, Celestijnenlaan 200D, 3001 Leuven, Belgium
     \and
     TU Graz, Fakult\"at f\"ur Mathematik, Physik und Geod\"asie, Petersgasse 16, 8010 Graz, Austria
            \and 
        Department of Chemistry and Molecular Biology, University of Gothenburg, Kemig\aa rden 4, 412 96 Gothenburg, Sweden
     }

\date{Received --------; accepted ---------}

% \abstract{}{}{}{}{} 
% 5 {} token are mandatory
 
  \abstract
  % context heading (optional)
  % {} leave it empty if necessary  
   {Clouds seem unavoidable in cool and dense environments, and hence, are necessary to explain observations of exoplanet atmospheres, most recently of WASP 96b 
   with JWST. 
   Understanding the formation of cloud condensation nuclei in non-terrestrial environments
   is therefore crucial to develop accurate models to interpret  present and future observations.}
  % aims heading (mandatory)
   {\red{The goal of the paper is to} support observations with infrared spectra
   for (TiO$_2$)$_N$ clusters in order to study cloud formation  in exoplanet atmospheres.}
  % methods heading (mandatory)
   {Vibrational frequencies  are derived from quantum-chemical calculations for 123 
   (TiO$_2$)-clusters and their isomers, and 
    line-broadening mechanisms 
   are evaluated. 
   Cluster spectra are calculated for several atmospheric levels
   for two example exoplanet atmospheres (WASP 121b-like and WASP 96b-like)
   to  identify possible spectral fingerprints for cloud formation.}
  % results heading (mandatory)
   { Rotational motion of and transitions in the clusters
   cause significant line broadening, so that individual vibrational lines are broadened beyond the spectral resolution of the medium resolution mode of the JWST mid-infrared instrument MIRI at R = 3000.  However, each individual cluster isomer exhibits a ``fingerprint'' IR spectrum. In particular, larger (TiO$_2$)-clusters have distinctly different spectra from smaller clusters. Morning and evening terminator for the same planet can exhibit different total absorbances due to different cluster sizes being more abundant.} 
  % conclusions heading (optional), leave it empty if necessary 
   {The largest (TiO$_2$)-clusters are not necessarily the most abundant (TiO$_2$)-clusters in the high-altitude regions of ultra-hot Jupiters, and the different cluster isomers will contribute to the local absorbance.  Planets with a considerable day-night asymmetry will be most suitable to search for  (TiO$_2$)-cluster isomers in order to improve cloud formation modelling.}
   

   \maketitle
%
%-------------------------------------------------------------------

\section{Introduction}

The presence of clouds has been inferred for many exoplanet atmospheres, as an explanation for suppressed absorption features and overall flat spectra \citep{Charbonneau2002DetectionAtmosphere,Pont2008DetectionTelescope,Kreidberg2014Clouds1214b}. For hot Jupiters, 
cloud models predict that cloud particles are primarily made of a mix of silicates and other metal oxides \citep{Helling2008DustLayers,Molliere2017ObservingObservations}. Silicate spectra exhibit a vibrational Si-O stretching feature at $10\mu m$ \citep{Hackwell1970InterstellarBands}. 
 With the launch of JWST in December 2021 and its near infrared observation capabilities in NIRCam, NIRSpec, and MIRI, the near infrared wavelength range ($\lambda = 0.6-5 \mu m$ for NIRCam and NIRSpec and $\lambda = 5-28 \mu m$ for MIRI) became accessible for observations at medium spectral resolutions. \cite{Miles2022TheB} provided the first direct detection of absorption around the $10\mu m$ band and therefore of silicate-based clouds in a planetary mass companion, observing VHS 1256b with JWSTs NIRSpec and MIRI MRS instruments. Observations released for the JWST initial science release target WASP 96b showed that clouds were needed to explain its spectral features. Additionally models for future JWST targets, such as WASP 121b,  or WASP 43b, exist and predict their cloud coverage or lack thereof \citep{Helling2021CloudWASP-18b}.  
 In order to quantify the impact of clouds on the chemistry and elemental abundances in exoplanet atmospheres, their formation processes need to be understood. Cloud particles condense from supersaturated gas on so-called cloud condensation nuclei (CCN). On gaseous exoplanets, these CCN need to be formed from the gas phase through a process called nucleation. This process is the first step of cloud formation and starts 
 as soon as the densities of the nucleating species are high, and the temperatures low enough. During the nucleation process differently sized small clusters or different isomers of the same sized cluster of the CCN-forming species co-exist 
 and it is unknown if only one or all  participate in the chemical path to a thermally stable CCN which then triggers the formation of a cloud particle.
 These clusters may have distinct spectral fingerprints, different to both that of the monomer and the bulk. Finding such spectral fingerprints may put the cloud formation modelling for extrasolar planets on much firmer grounds. The aim of this paper is therefore to provide harmonic vibrational spectra of these small clusters for the nucleating species Titanium dioxide (TiO$_2$) to  enable discoveries similar to the unexpected SO$_2$ detection in WASP-39b with JWST \citep{Rustamkulov2023JWSTPRISM}. These spectra arise from excited vibrational states of the clusters and their isomers that have energies that correspond to wavelengths in the 
infra-red 
range that
can be probed by JWSTs MIRI. Rotational 
transitions have energies in the millimetre wavelength ranges, but they also happen within vibrational state transitions, broadening the corresponding vibrational bands. 

Section \ref{sec:methods} describes how the harmonic vibrational spectra are extracted from quantum-chemical calculations performed in \cite{Sindel2022RevisitingEnvironments} and the broadening mechanisms considered. Section \ref{sec:Results} compares the spectra for different clusters and isomers and presents the resulting total absorbance caused by (TiO$_2$)$_N$ clusters for two exoplanet atmosphere models. In Section \ref{sec:discusssion}, we discuss our results and their limits. Section \ref{sec:conclusions} summarises the paper and gives and outlook on possible future work.

%--------------------------------------------------------------------
\section{Methods}
\label{sec:methods}
\subsection{Extraction of harmonic IR frequencies}
\label{sec:extrac}
This work investigates the infrared spectra of small (TiO$_2$)$_N$ 
clusters.
The harmonic vibrational spectra discussed in this work are based on the results of density functional theory (DFT) frequency calculations performed with the B3LYP functional \citep{Becke1993Density-functionalExchange}, the cc-pVTZ basis-set \citep{Wilson1996GaussianNeon} and gd3bj empirical dispersion \citep{Grimme2011EffectTheory}. The cluster structure and energy calculations were presented in \cite{Sindel2022RevisitingEnvironments}, for (TiO$_2$)$_N$ 
clusters of sizes $N$ = 1-15. For all clusters with sizes $N > 1$, multiple isomers are considered. 
(see Table \ref{tab:isopercluster}). For each cluster size $N$, there is one most energetically favourable cluster 
configuration, the so-called global minimum candidate (GM).
\red{The GM is given the suffix -A, e.g. TI6O12-A denotes the GM cluster of size 6. The second lowest-energy isomer, i.e. the energetically next favourable cluster at T = 0K is given the suffix -B, the third -C, etc..}

\begin{table}            % title of Table   % is used to refer this table in the text
\centering          
\caption{Number of isomers for each cluster size $N$}
\label{tab:isopercluster}  % used for centering table
\begin{tabular}{|c|c|}        % centered columns (4 columns)
\hline                 % inserts double horizontal lines
$N$ & \# of isomers \\    % table heading 
\hline                        % inserts single horizontal line
1 & 1 \\
2 & 3 \\
3 & 7 \\
4 & 4 \\
5 & 4 \\
6 & 6 \\
7 & 7 \\
8 & 10 \\
9 & 7 \\
10 & 12 \\
11 & 14 \\
12 & 13 \\
13 & 14 \\
14 & 10 \\
15 & 11 \\
\hline    
Total & 123 \\
\hline%inserts single line
\end{tabular}
\end{table}

The frequencies and their corresponding infrared (IR) intensities are given in units of [cm$^{-1}$] and $\frac{\rm km}{\rm mol}$, respectively. We note that some of the low energy / high wavelength modes ($\lambda \gtrsim 77 \mu m$) are not pure vibrational modes, but can arise from hindered rotation, where certain rotations of the molecule are inhibited due to its geometry. In this work, we treat all modes as purely harmonic vibrational modes. We convert the IR intensities to molar absorption coefficients according to \cite{Spanget-Larsen2015IROutput}.  The unknown quantity for this 
transformation is the line-width for each active IR mode. To quantify the line-width several broadening 
mechanisms are evaluated for the physical parameterspaces of exoplanetary atmospheres: 

\subsubsection{Thermal broadening}
Thermal broadening of a line is caused by the multi-directional movement of individual line-sources within a gas. The broadening in frequency ($\Delta \nu$) by thermal broadening is dependent on the temperature ($T$), the frequency ($\nu$) and the mass of the individual line-sources, i.e. the cluster, ($m$) through:
\begin{equation}
\label{eq:thermal_broadening}
    \frac{\Delta \nu}{\nu} = \frac{1}{c} \sqrt{\frac{2 k_B T}{m}}
\end{equation}
where $k_B$ is the Boltzmann constant and $c$ is the speed of light.
As a consequence, large and more massive clusters show smaller thermal line broadening than small clusters.
Here, we assume that each active frequency mode of a cluster is broadened equally.

\subsubsection{Collisional broadening}
In a dense gas, usually spontaneous emissions can be induced, i.e. triggered prematurely, through collisions with other particles. This shortens the average lifetime of the state and therefore increases the uncertainty in energy and therefore frequency. The relative broadening at a frequency is given by
\begin{equation}
    \frac{\Delta \nu}{\nu} = \frac{2 \nu_{col}}{\nu}
\end{equation}
where $\nu_{col}$ is the frequency of collisions within the gas, which is given by the root mean square (RMS) velocity of the gas particles ($v_{rms}$) and the mean free path ($\lambda$):
\begin{equation}
    \nu_{col} = \frac{v_{rms}}{\lambda}
\end{equation}
with the RMS velocity of a gas being:
\begin{equation}
    v_{rms} = \sqrt{\frac{3RT}{M}},
\end{equation}
with the ideal gas constant $R$, the gas temperature $T$ and the mean molecular weight of the gas $M$.
The mean free path is given by:
\begin{equation}
    \lambda = \frac{RT}{\sqrt{2}\pi \sigma^2 N_A p},
\end{equation}
with the gas pressure $p$, the effective cross-section of the cluster $\sigma$ and the Avogadro constant $N_A$.

\subsubsection{Rotational broadening}
A molecule with $n$ atoms has $3n-6$ vibrational modes. The rotational motion of the molecule can affect the vibrational motion and vice versa. This means that the vibrational energy levels are slightly dependent on the rotational energy levels. As a result, when rotational transitions occur, they can slightly shift the vibrational energy levels, which then leads to additional broadening of the vibrational spectral lines. The population of rotational states J follows a Boltzmann distribution. In order to quantify the broadening mechanism induced by these rotational states, the maximum rotational quantum number
$J_{max}$ is used. It is defined by:
\begin{equation}
    J_{max} = \left(\frac{k_B T}{2hcB}\right)^{\frac{1}{2}} - \frac{1}{2}.
\end{equation}
$B$ is the rotational constant of the molecule and defines the spacing of rotational energy levels. This maximum rotational quantum number specifies the most likely occupied rotational energy state \citep{Hollas2002BasicSpectroscopy}. Since the broadening effect is symmetric, the frequency space spanned by the rotational states from $J=0$ to $J_{max}$ is taken as the FWHM for our broadening approximation.
The broadening induced by this process is therefore given as:
\begin{equation}
    \frac{\Delta \nu}{\nu} = \frac{2BJ_{max}}{\nu}
\end{equation}

\subsubsection{Natural broadening}
Natural broadening occurs due to the Heisenberg uncertainty principle. An uncertainty in the life-time of an excited state ($\Delta t$) leads to an uncertainty in its energy ($\Delta E$) and therefore in in its frequency ($\Delta \nu$).
\begin{equation}
\label{eq:natbroad}
    \Delta \nu \approx \frac{\Delta E}{h} \approx \frac{1}{2 \pi \Delta t}
\end{equation}
Natural broadening effects are usually small when compared to other broadening effects (Fig. \ref{fig:broadening_comp}) so they are not considered in this work. 

\subsection{Number densities of (TiO$_2$)$_N$ clusters}
In order to calculate not only the relative molar absorbances of the individual (TiO$_2$)$_N$ clusters at different pressure and temperature points, but the absolute absorbance of each species at certain pressure levels in the atmosphere, number densities for
as many
species 
as possible
need to be included, because they all contribute differently to the overall absorption. In order to achieve that, we input the
thermochemical cluster-data into the equilibrium chemistry code {\sc GGChem} \cite{Woitke2018EquilibriumRatio}.
The molecular equilibrium constants are fit using the approach by \cite{Stock2018FastChem:Atmospheres}. The Gibbs free energies of formation ($\Delta G^\circ_f$) are calculated by
\begin{equation}
\Delta G^\circ_f(({\rm TiO}_2)_N) = \Delta G_{DFT}^\circ({\rm TiO})_2)_N - N \Delta G_{Janaf}^\circ({\rm Ti}) - 2N \Delta G_{Janaf}^\circ ({\rm O}).
\end{equation}
Values for $\Delta G_{DFT}^\circ({\rm TiO}_2)_N$ are taken from \cite{Sindel2022RevisitingEnvironments}, while $\Delta G^\circ_{Janaf}$ for Ti and O are taken from the JANAF-NIST thermochemical tables \citep{MalcolmW.Chase1998NIST-JANAFTables}.
The resulting fit for the $k_p$ values was then done through Equations (13) and (17) in \cite{Woitke2018EquilibriumRatio} which combine into:
\begin{equation}
    \Delta G^\circ_f (T) = -RT \left( \frac{a_0}{T} + a_1 \ln{T} + b_0 + b_1T + b_2T^2 \right)
\end{equation}
with the fit-coefficients $a_0$, $a_1$, $b_0$, $b_1$, and $b_2$. These were then implemented into {\sc GGChem} for 
the 123 considered (TiO$_2$)$_N$ cluster isomers (Tab. \ref{tab:isopercluster}), enabling the calculation of their number densities from equilibrium chemistry at any pressure temperature point with $T > 100$K. 

\subsection{Atmospheric pressure - temperature profiles}
\label{sec:pt_prof}
We use 1D pressure-temperature ($p_{\rm gas}$, $T_{\rm gas}$)-profiles that were extracted from \cite{Baeyens2021GridMixing} and extended to pressures of  $10^{-15}$ bar. We focus on two specific cases: A WASP-121b like planet ($T_* = 6500$K, $T_{\rm eff} = 
2200$K, log(g) = 3) and a WASP-96b like planet ($T_* = 5650$K, $T_{\rm eff} = 1200$K, log(g) = 3). WASP 96b is a puffy hot Jupiter, that is cold enough to produce cloud coverage everywhere but the substellar point, and the existence of clouds has been 
predicted by cloud formation models \citep{Samra2023CloudsWASP-96b}. WASP 121b is an ultra-hot Jupiter with extremely high day/nightside temperature differences, where clouds are only expected to form on the nightside \citep{Helling2021CloudWASP-18b}. For both of the planets a ($p_{\rm gas}$, $T_{\rm gas}$)-profile for the morning and the evening terminator are used.

\subsection{Absorbance of clusters}
For all of the ($p_{\rm gas}$, $T_{\rm gas}$)-profiles, we start from a gas of solar composition \citep{Asplund2009TheSun} and run {\sc GGChem} for gas-phase species only at every point along the ($p_{\rm gas}$, $T_{\rm gas}$)-curve. This provides the number densities for all isomers. At each pressure and temperature point, all broadening mechanisms are evaluated for all vibrational lines of all clusters and the strongest broadening mechanism is chosen. \red{The broadening through this method is used as the input broadening width $w$} to convert the lines IR intensity into an absorption profile $\sigma$ with units of [cm$^2$/molecule].
All absorption profiles are mutliplied by the number density [$\frac{\rm molecule}{\rm cm^3}$] of their respective cluster, giving their total absorbance $A_{tot}$ in [$\frac{1}{\rm cm}$]:
\begin{equation}
    A_{tot, cluster} = n_{\rm cluster} \sigma_{\rm cluster}
\end{equation}

%-----------------------------------------------------------------
\section{Results}
\label{sec:Results}
\subsection{Broadening mechanisms}
\label{sec:results_broadening}
\begin{figure}
    \centering
    \includegraphics[width = \columnwidth]{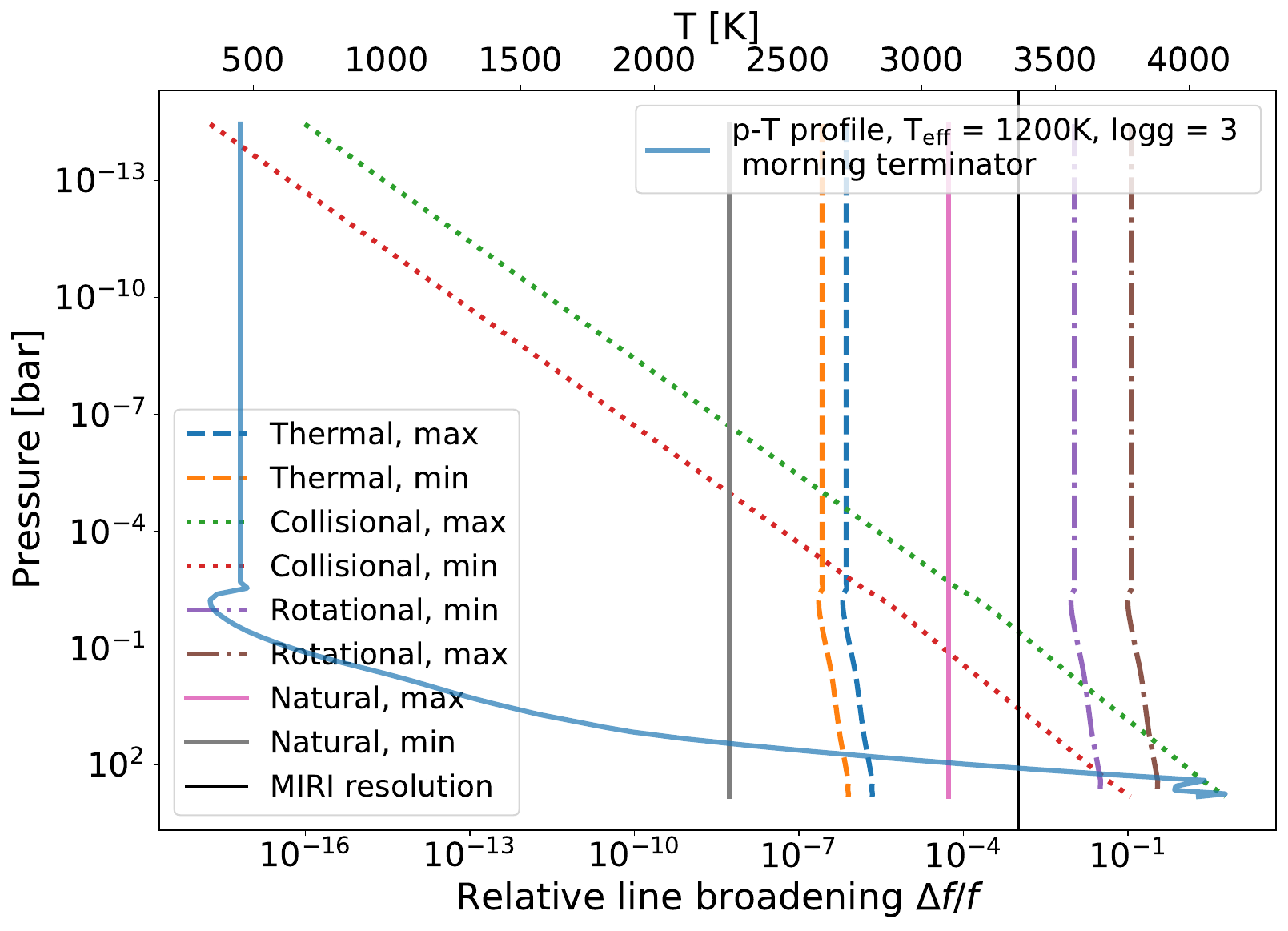}
    \caption{Comparison of the considered broadening mechanisms along a ($p_{\rm gas}$, $T_{\rm gas}$)-profile. All broadening mechanisms are evaluated for their potential minimum and maximum contribution. \red{Thermal broadening is weakest for large clusters ($N = 15$, 'Thermal, min'), and strongest for small clusters ($N = 1$, 'Thermal, max'). Collisional broadening is weakest for small clusters at high frequencies ($N=1$, $\lambda = 10\mu m$, 'Collisional, min') and strongest for large clusters at low frequencies ($N=15$, $\lambda = 100\mu m$, 'Collisional, max'). Rotational broadening is weakest for large clusters ($N=15$, 'Rotational, min') and strongest for small clusters ($N=1$, 'Rotational, max'). Natural broadening is weakest for long state life-times at high frequency ($\tau = 1\mu s$, $\lambda = 10\mu m$, 'Natural, min') and strongest for short state life-times at low frequency ($\tau = 1ns$, $\lambda = 100 \mu m $, 'Natural, max'). The T(p)-profile is one of the WASP-96b-like profiles from Sec. \ref{sec:pt_prof}.}
    }
    \label{fig:broadening_comp}
\end{figure}

The impact of the broadening mechanisms at different pressure levels at the morning side terminator of a WASP-96b like planet is investigated as an example (Figure \ref{fig:broadening_comp}). 
It is of interest whether the individual vibrational modes could be resolved by JWSTs MIRI instrument at its working spectral resolution of $R \sim 3000$. Any mechanisms that broaden the lines further than that will make them resolvable at this resolution. 

The impact of collisional broadening depends strongly on the pressure. In our test case atmosphere it is only relevant at high pressures low in the atmosphere. For the atmosphere used in this test case, the upper layers are largely isothermal. Therefore broadening mechanisms that depend predominantly on temperature such as thermal and rotational broadening are constant throughout most of the atmosphere and only change when the temperature changes too. For natural broadening, the smallest broadening effects are characterised by a long state life-time  and a small emission wavelength (Eq. \ref{eq:natbroad}). $\tau = 1\mu s$ and $\lambda = 10\mu m$ are chosen representatively.  The maximum broadening scenario corresponds to a short state life-time ($\tau = 1ns$) and a large emission wavelength ($\lambda = 100\mu m$). In both cases, the relative line broadening is far below the rotational broadening. Therefore, it is justified to neglect natural broadening.
The most impactful broadening mechanism throughout most of the atmosphere is rotational broadening, dominating everywhere but deep in the atmosphere.

\subsection{Raw spectra of (TiO$_2$)$_N$ clusters}
\label{sec:results_rawspec}
\begin{figure}
    \centering
    \includegraphics[width = \columnwidth]{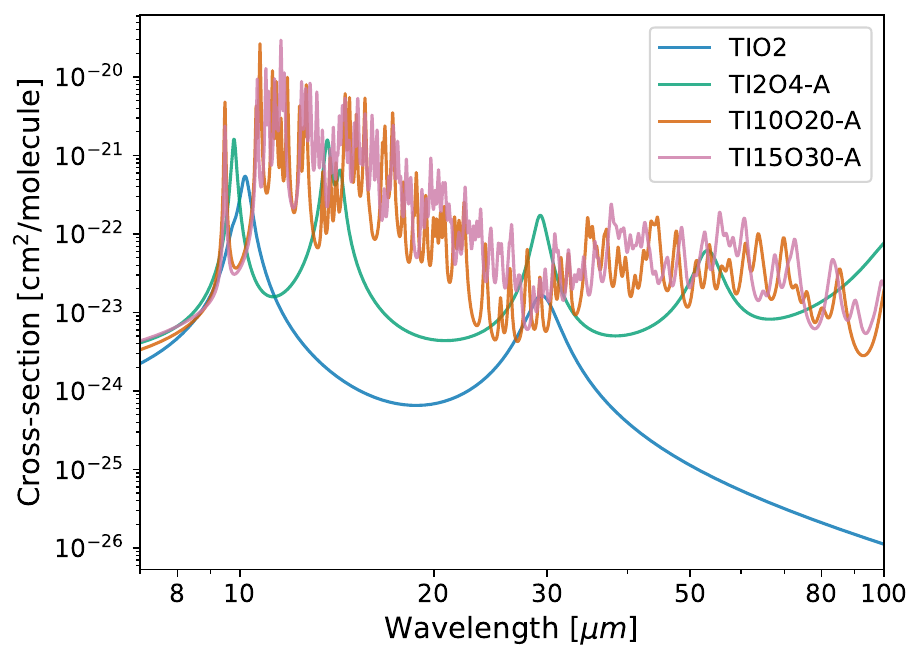}
    \caption{Wavelength dependent cross-sections $\sigma_{\rm cluster} (\lambda)$ for the GM clusters of sizes $N = 1,2,10,15$.}.
    \label{fig:spec_comparison_size}
\end{figure}

Every (TiO$_2$)$_N$ cluster contains $X=3N$ atoms and shows $3X-6$ vibrational modes. Therefore,  the overall opacity is expected to increase with the cluster size $N$.
Wavelength dependent cross-sections $\sigma_{\rm cluster} (\lambda)$ [$\frac{\rm cm^2}{\rm molecule}$] are computed for 123 isomers of sizes $N = 1-15$ of the (TiO$_2$)$_N$ clusters. The broadening for each line at a temperature of $T = 1000{\rm K}$ and a pressure of $p = 1$ bar is calculated and the lines are broadened according to \cite{Spanget-Larsen2015IROutput}:
Each line profile is given by a Lorentzian with a FWHM of $w$ and a maximum of $\epsilon_{\rm max} = 27.648 \frac{I_{\rm IR}}{w}$. \red{The width $w$ is computed by comparing the broadening mechanisms discussed in Sec. \ref{sec:extrac} and using the width produced by the dominant one.}
A comparison is made to show trends between different sizes in Figure \ref{fig:spec_comparison_size}. 
In the whole spectral range of $\lambda = 5 - 100\mu m$ cross-sections of larger clusters tend to be larger than ones of smaller clusters. 
Especially in the wavelength-range observed by the JWST/MIRI instrument of $5-28 \mu m$, larger clusters are 
stronger absorbers than smaller clusters, while the TiO$_2$ monomer has no significant absorption cross-section over 
most of this range. 
Cluster growth (i.e. nucleation) will therefore lead to an increase of absorption in the 
$\sim 4-25 \mu m$ wavelength region, with the 
largest clusters 
dominating the absorption. 

\begin{figure}
    \centering
    \includegraphics[width = \columnwidth]{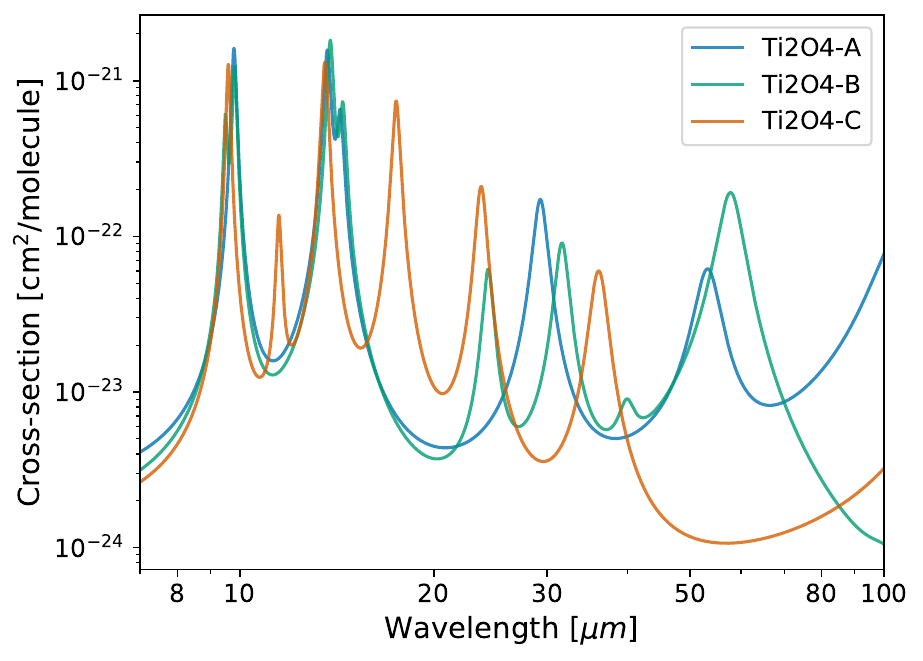}
    \caption{Wavelength dependent cross-sections $\sigma_{\rm cluster} (\lambda)$ for all 3 isomers of the (TiO$_2$)$_2$ cluster.}.
    \label{fig:spec_comparison_iso_2}
\end{figure}

\begin{figure}
    \centering
    \includegraphics[width = \columnwidth]{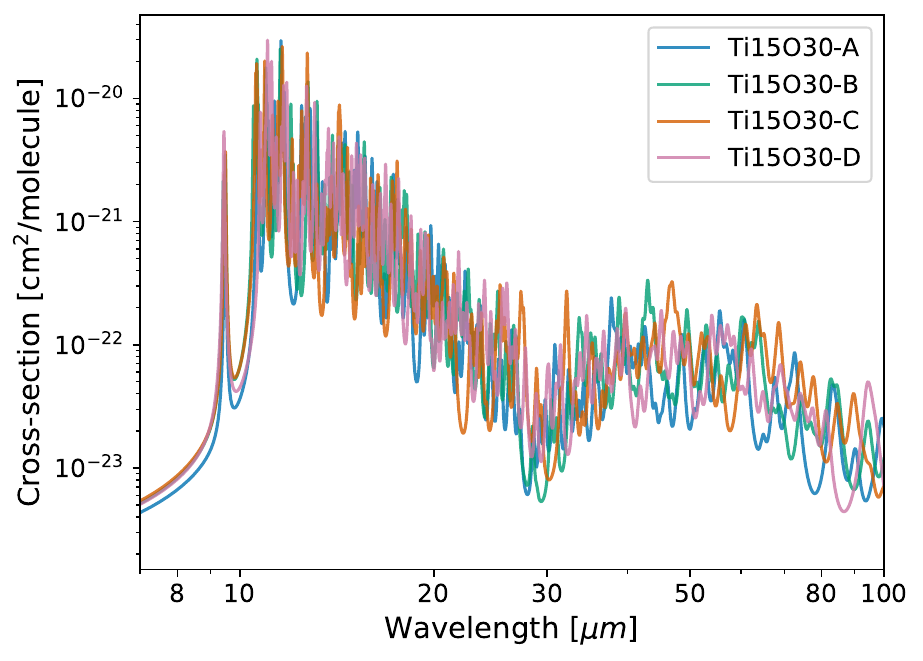}
    \caption{Wavelength dependent cross-sections $\sigma_{\rm cluster} (\lambda)$ for the 4 energetically most favorable isomers of the (TiO$_2$)$_{15}$ cluster.}.
    \label{fig:spec_comparison_iso_15}
\end{figure}

Generally, not only the GM cluster will be present, but energetically less favourable isomers will also be present in the atmosphere. Since their geometries are different, their vibrational spectra also differ. A comparison is made of cross-sections for 
the 3 isomers of size $N=2$ (Fig. \ref{fig:spec_comparison_iso_2}) and the 4 
lowest-energy isomers of size $N=15$ (Fig. \ref{fig:spec_comparison_iso_15}).
    
For the molecular cross-sections of a small 
cluster, (TiO$_2$)$_2$, there are differences between the isomers, especially for wavelengths towards the tail at $100 \mu m$. Within the MIRI wavelength range ($5-28 \mu m$) they agree in key features, such as the peak just short of $\sim 10 \mu m$ and the double peak at $\sim 15 \mu m$ for the 2 
most favourable
isomers. For the large 
cluster (TiO$_2$)$_{15}$, the forest of lines at shorter wavelengths, covering most of the MIRI wavelength range can only be disentangled at high spectral resolutions. At longer wavelengths differences in line positions become more clear. It is important to note that all 4 energetically favorable isomers share a strong line feature at $\sim 10 \mu m$. 

\subsection{Cluster number densities in chemical equilibrium}
\begin{figure}
    \centering
    \includegraphics[width = \columnwidth]{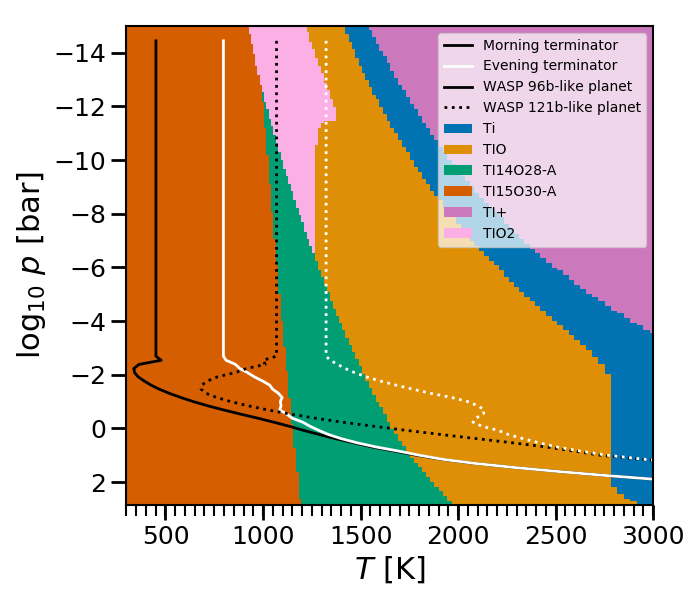}
    \caption{Most abundant Ti-containing species
    in chemical equilibrium at solar metallicity as a function of temperature and pressure. Letters after clusters indicate \red{energetic} isomer ordering within isomers of the same size, i.e. -A indicates the GM isomer, \red{while -B would indicate the second most energetically favourable isomer}. Overplotted are temperature pressure profiles for morning (in black) and evening (in white) terminators of a WASP 96b-like planet and a WASP 121b-like extrasolar planet.}
    \label{fig:phasediagram_ti_ptprof}
\end{figure}

\begin{figure}
    \centering
    \includegraphics[width = \columnwidth]{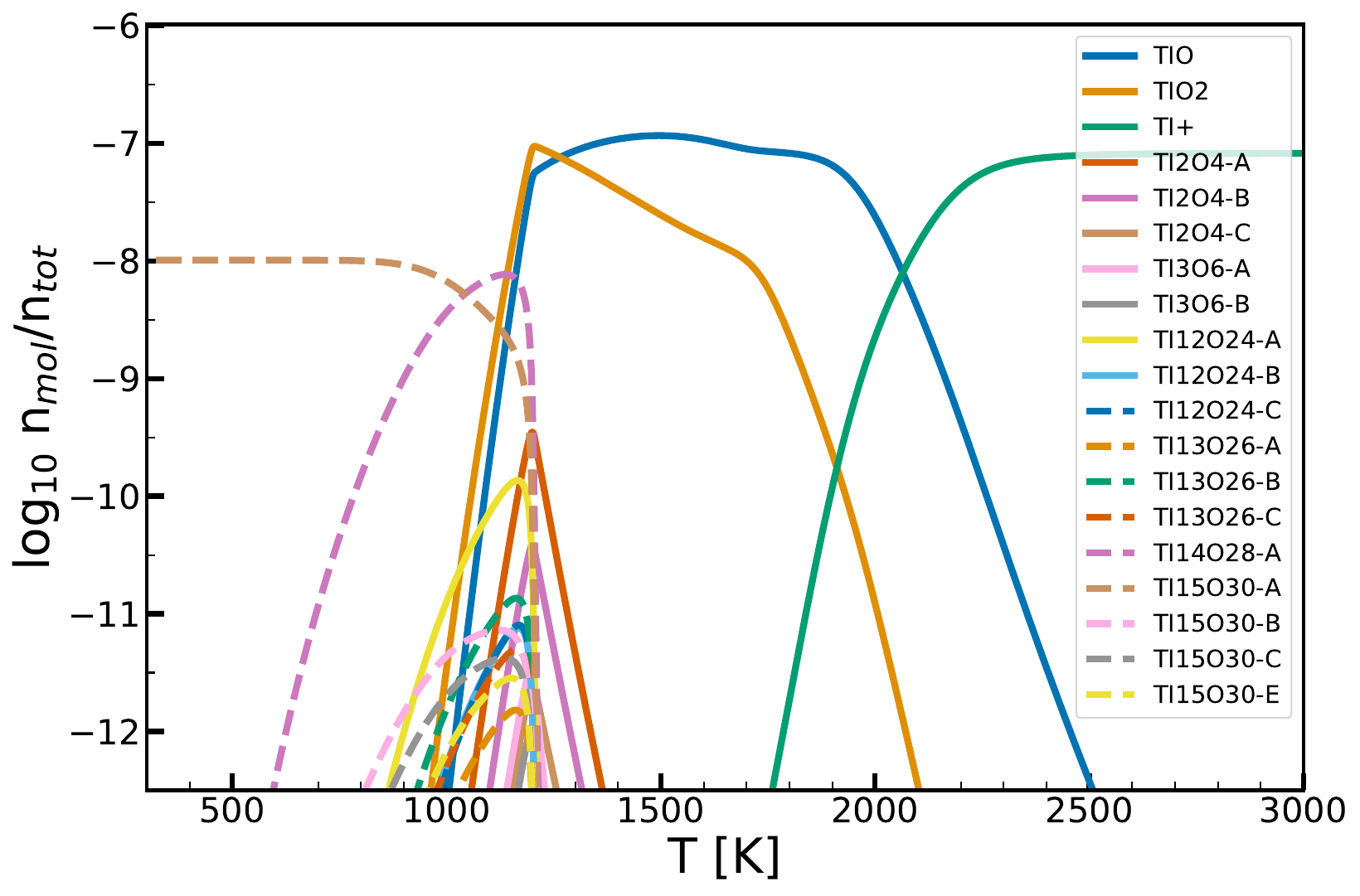}
    \caption{Concentration of Ti-bearing gas-phase species across a temperature range of $300 - 3000$K. \red{Letters after cluster names indicate different isomers, with -A being the energetically most favourable (GM) isomer.} Only species that reach a concentration of $\frac{n_{mol}}{n_{tot}}>10^{-12}$ are shown.}
    \label{fig:1d_1e-8bar}
\end{figure}

In order to calculate the total absorbance of a species the cross-sections for each isomer of each cluster need to be multiplied by its respective number density in the atmosphere. The chemical equilibrium code {\sc GGChem} is used to compute the number densities of all Ti-bearing molecules and 
clusters from solar metallicity in a grid of pressures from $p = 10^{-15} - 700$ bar and temperatures of $T = 300 - 3000$K (Fig. \ref{fig:phasediagram_ti_ptprof}).
On the WASP 96b-like planet (solid lines) the cluster composition
in the upper atmosphere is dominated by the largest cluster available in the calculations (TiO$_2$)$_{15}$ for both terminators. It is reasonable to assume that larger clusters would be more favourable if included in the calculations. In the lower atmosphere between $\sim 1 - 100$bar, the largest cluster is no longer the most favoured Ti-bearing 
compound
, but the second-largest cluster, (TiO$_2$)$_{14}$ is more abundant. At these pressure levels, a limit to the stability of larger clusters has been reached, making the smaller cluster more favourable. 
For the WASP 121b-like planet (dotted lines), the terminators probe different
conditions leading to different atmospheric compositions.
At the top of the atmosphere at the morning terminator, the TiO$_2$ monomer is dominating, as any larger clusters are less 
favourable
at these temperatures and pressures. This changes when moving downwards in the atmosphere, increasing the pressure and therefore also the densities of the atmosphere. At pressures between $\sim 10^{-11} - 10^{-6}$ bar, some growth of clusters is favoured, again leaving the second largest cluster (TiO$_2$)$_{14}$ as the most abundant one. As pressures further increase, i.e. deeper in the atmosphere, larger clusters, here represented by (TiO$_2$)$_{15}$, become the most 
favourable
species ($p_{\rm gas} \approx 10^{-6} - 10^{-1}$ bar), after which the temperature increases beyond the stability limit and 
follows a similar path to the terminators of the WASP 96b-like planet. The hotter evening terminator starts within the TiO-dominated regime, where temperatures are high enough for the smaller molecule to be favoured over the TiO$_2$ monomer. This dominance continues down to $p \sim 10^{-5}$ bar, interrupted by a small section ($\sim 10^{-13} - 10{^{-11}}$ bar) where TiO$_2$ is most abundant. Larger clusters ($N\geq 15$) are never favoured along this terminator, only the (TiO$_2$)$_{14}$ cluster is most abundant between $p \approx 10^{-5} - 10^{-2}$ bar. Deep in the atmosphere, where temperatures are higher,
the 
titanium content
is dominated by first the more thermally stable TiO and at higher pressures the atomic Ti. A slice through this graph at $p=10^{-8}$~bar showcases the prevalence of different clusters at different temperatures (Fig. \ref{fig:1d_1e-8bar}).
For temperatures $T < 600$K the GM candidate of the largest cluster, (TiO$_2$)$_{15}$, is the only relevant Ti-bearing 
species. As the temperature rises, the second largest cluster (TiO$_2$)$_{14}$ rises in abundance, indicating a 
comparatively enhanced preference at certain temperatures and pressures,
as it becomes the most abundant species at $T \approx 1100$K. In the region between $\sim 900 - 1300$K low abundances of many intermediate sized GM clusters and their respective metastable isomers can be seen. Especially the size $N=12$ cluster (solid gold) and the $N=2$ cluster (solid red) play a role.
\red{The relative energies between isomers of the same size vary with temperature, so that different isomers may become the most abundant of their size at different temperature points. For $N=15$, the energy difference of isomers A and B at 300K is $\Delta(\Delta G_f^\circ(N=15, T=300{\rm K}, {\rm A-B}))) = -92.9 \frac{\rm kJ}{\rm mol}$. At 1500K, this difference decreases to $\Delta(\Delta G_f^\circ(N=15, T=1500{\rm K}, {\rm A-B}))) = -37.8 \frac{\rm kJ}{\rm mol}$. At 3000K, isomer B is more energetically favourable, as the difference is $\Delta(\Delta G_f^\circ(N=15, T=3000{\rm K}, {\rm A-B}))) = +31.5 \frac{\rm kJ}{\rm mol}$. The first isomer of $N=15$ that does not reach the cut-off concentration of $\frac{n_{mol}}{n_{tot}}>10^{-12}$ is TI15O30-D. The relative energy to the most stable $N=15$ isomer is always $\Delta(\Delta G_f^\circ) > -92.9\frac{\rm kJ}{\rm mol}$ across the entire temperature range 300-3000K. The relative cut-off energy is therefore at around $90\frac{\rm kJ}{\rm mol}$ for $N=15$.}

As the temperature increases beyond the stability for clusters, the abundance of the TiO$_2$ monomer rises. The difference in peak abundance between TiO$_2$ and (TiO$_2$)$_{15}$ is about an order of magnitude, 
indicating that a similar amount of Ti and O is used to produce these concentrations.

\subsection{Spectra at different atmospheric pressure levels}
To investigate the change of the absorption profiles of (TiO$_2$)$_N$ 
clusters at different pressure levels in the atmosphere, the spectra are computed at even logarithmic steps from $p = 10^{-15} - 10^2$~bar. At each pressure level, {\sc GGChem} is run to obtain the number densities of all considered species, i.e. the 576 atoms and molecules included in {\sc GGChem} as well as the 123 clusters from this work. As the molecular cross-sections span no more than 5 orders of magnitude, contributions from clusters with abundances more than 5 magnitudes lower than the most abundant cluster are neglected to save
computational cost for the absorbance calculation. For each of the remaining species, the absorbance is calculated by multiplying the molecular cross-section with its number density. All  absorbances are then added up to the total absorbance of (TiO$_2$)$_N$ clusters at the given pressure level. 
\begin{figure}
    \centering
    \includegraphics[width = \columnwidth]{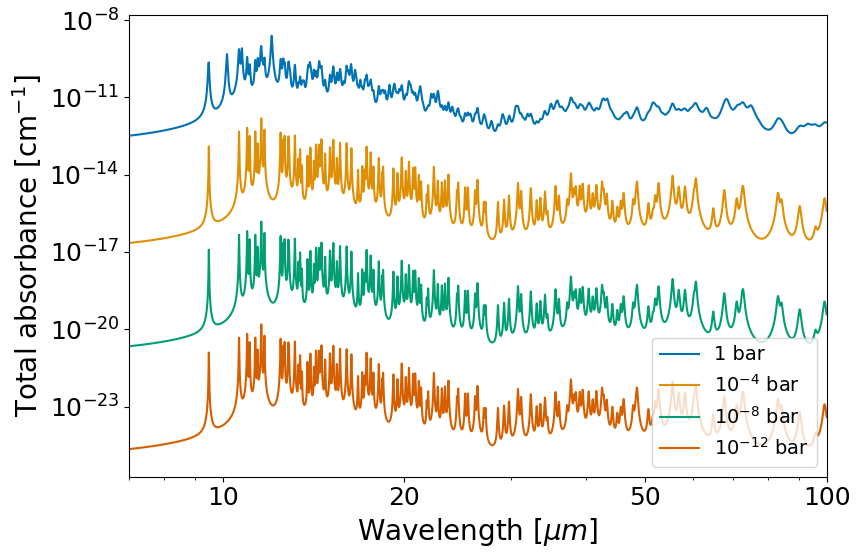}
    \includegraphics[width = \columnwidth]{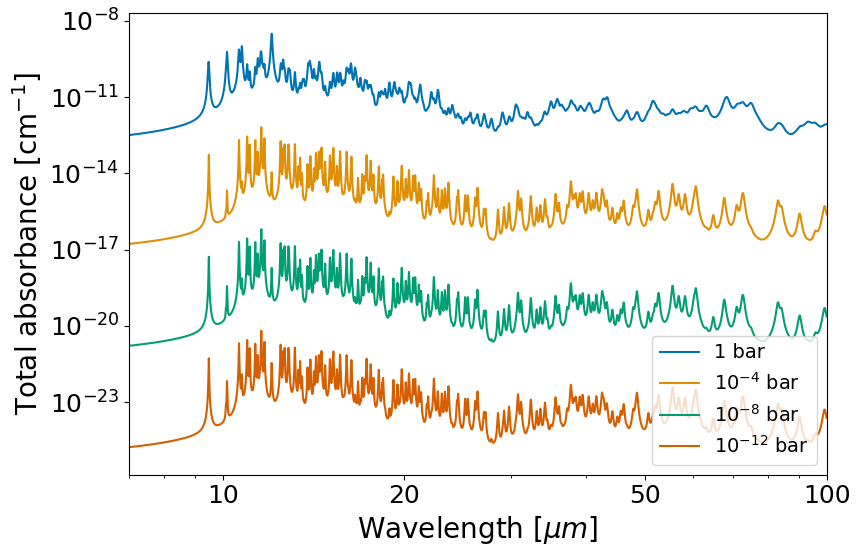}
    \caption{Total absorbance from (TiO$_2$)$_N$ 
    clusters at different pressure levels of a WASP 96b-like exoplanet atmosphere as a function of the wavelength. \textbf{Top:} Morning terminator. \textbf{Bottom:} Evening terminator.}
    \label{fig:spec_wasp96b_miri}
\end{figure}

For the WASP 96b-like planet (Fig. \ref{fig:spec_wasp96b_miri}) there are only minor differences between morning and evening terminator, owing to the fact that both temperature-pressure profile occupy similar regions in the phase-diagram (Fig. \ref{fig:phasediagram_ti_ptprof}). At the bottom of the atmosphere at 1 bar, there are many cluster-sizes and their isomers contributing to the total spectrum, blending lines and causing the lack of distinct peaks towards the longer-wavelength part of the spectrum. Additionally the higher pressure causes more collisional broadening, further blending lines. This changes when moving upwards in the atmosphere, where the 
Ti content
is dominated by the large cluster (TiO$_2$)$_{15}$, with its distinct lines across the wavelength range. The most visible difference between the morning and evening terminators are the small peaks at $\sim 10$ and $12 \mu m$, that exist for the evening terminator but not for the morning terminator. As the evening terminator is hotter, 
(TiO$_2$)$_{15}$ becomes less favourable
and some contribution from the next smaller cluster is visible, as these peaks are caused by absorption from (TiO$_2$)$_{14}$. 

An overall trend is that the absorbance is higher at higher pressures, i.e. lower in the atmosphere. This is due to the fact that number densities are much higher at the bottom of the atmosphere.
\begin{figure}
    \centering
    \includegraphics[width = \columnwidth]{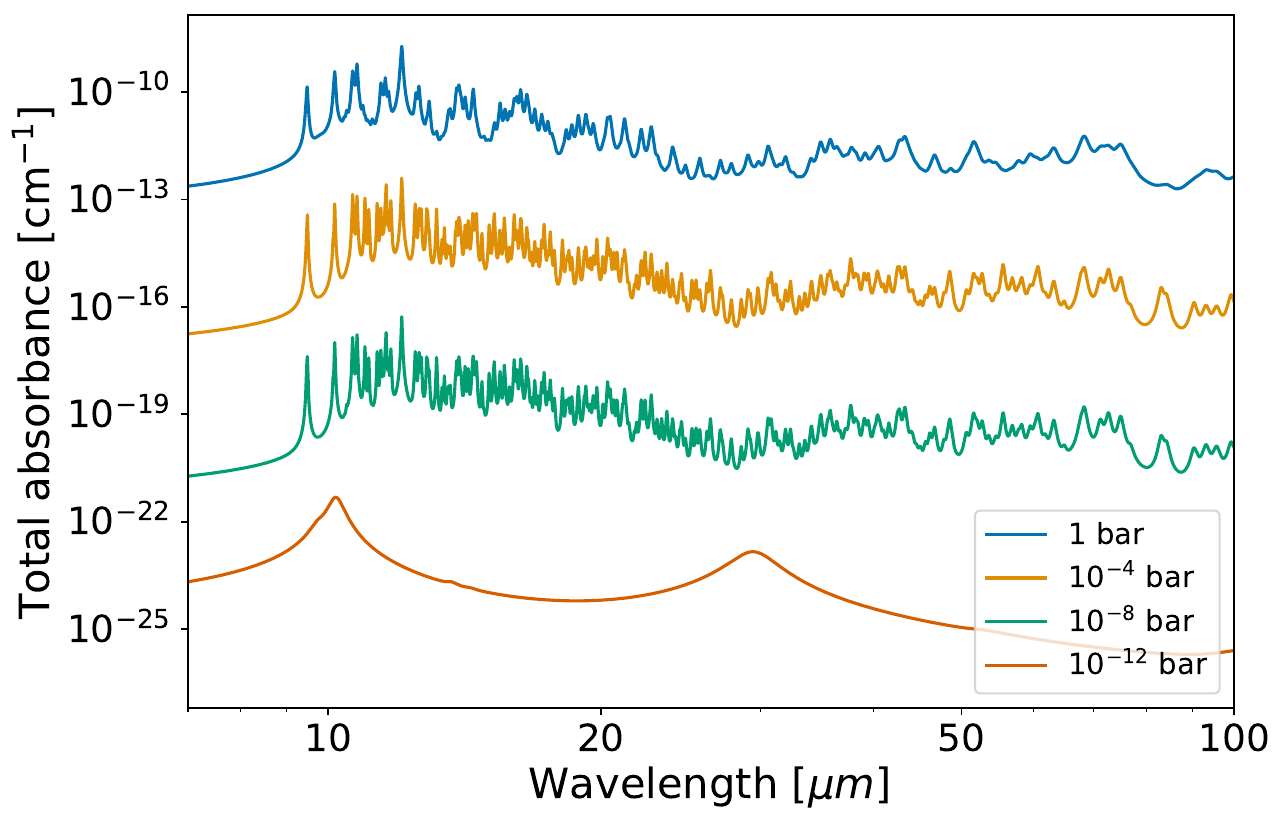}
    \includegraphics[width = \columnwidth]{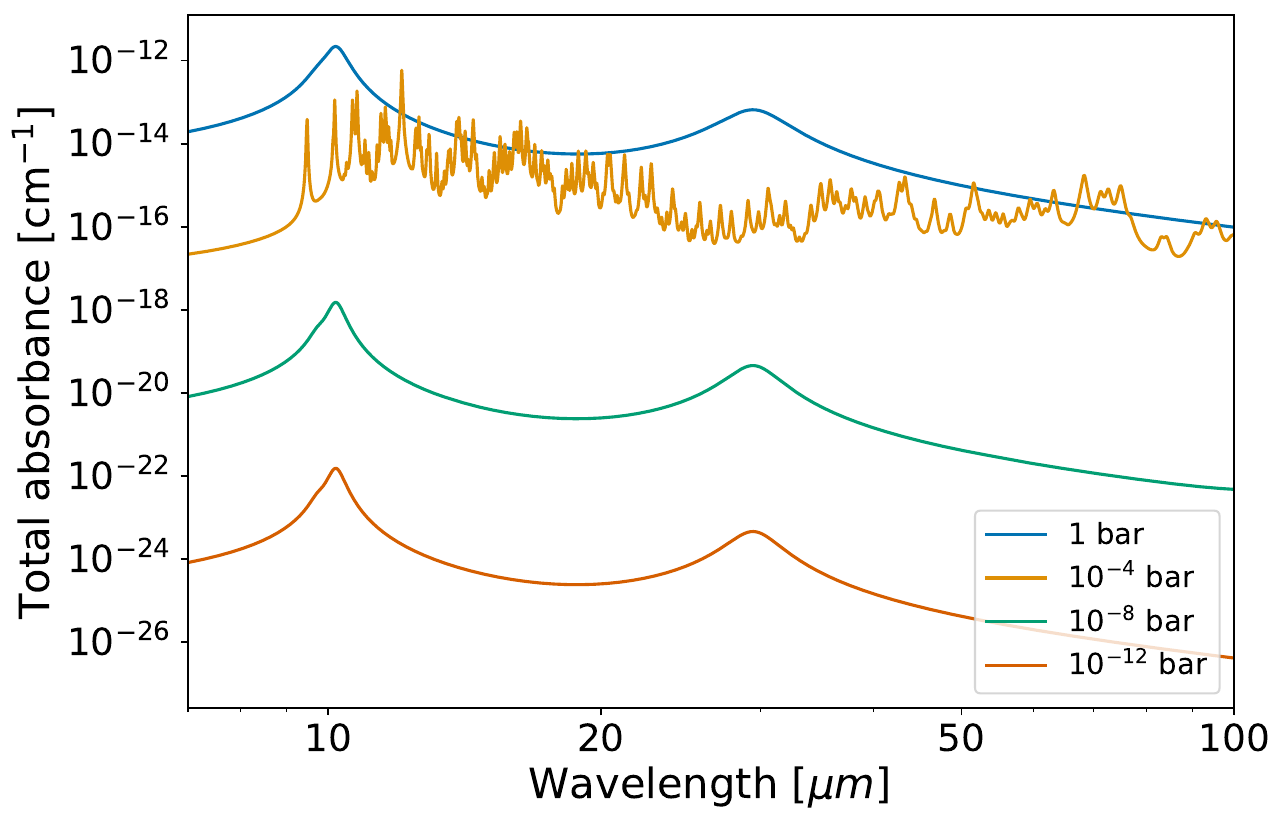}
    \caption{Total absorbance from (TiO$_2$)$_N$ 
    clusters at different pressure levels of a WASP 121b-like exoplanet atmosphere as a function of the wavelength. \textbf{Top:} Morning terminator. \textbf{Bottom:} Evening terminator.}
    \label{fig:spec_wasp121b_miri}
\end{figure}

The spectra for the WASP 121b-like planet (Fig. \ref{fig:spec_wasp121b_miri}) are more easily distinguishable with respect to their terminators. At the evening terminator the only contributing TiO$_2$ species is the monomer for low pressures levels ($p = 10^{-12}, 10^{-8}$ bar) and high pressure levels ($p = 1$bar). The TiO molecule is the dominant Ti-bearing species for most high and low pressures ($p < 10^{-6}$ bar and $p > 10^{-2}$ bar).  Only at an intermediate pressure level ($p = 10^{-4}$ bar) 
clusters may contribute to the spectrum. As the second largest cluster (TiO$_2$)$_{14}$ is the most abundant here, the peaks at 10 and 12$\mu m$ are well pronounced. It also showcases the strength of absorption of larger clusters in comparison to the monomer, as the total absorbance with the contribution of the clusters approaches, and sometimes exceeds, the absorbance of 
the monomer at much higher pressures. For the morning terminator only the upper atmosphere is dominated by the TiO$_2$ monomer. Lower in the atmosphere the spectra have contributions from many clusters and isomers, as both (TiO$_2$)$_{14}$ and (TiO$_2$)$_{15}$ and its isomers are highly abundant. At the bottom of the atmosphere ($p = 1$ bar), fewer isomers and clusters are 
present
due to the higher temperature, leading to less distinct lines, as well as the higher collisional broadening causing the lines to blend into each other.

\section{Discussion}
\label{sec:discusssion}
The (TiO$_2$)$_N$ clusters 
considered in this work 
extend to a maximum size of $N=15$. 
As the largest cluster is the most abundant one over a large area of the parameter space (Fig. \ref{fig:phasediagram_ti_ptprof}), this has some implications for our results. When the largest considered cluster is also the most abundant one, larger clusters, which are more stable when extrapolating observed trends, 
are likely to be more abundant. 
It is therefore important to note that the spectra at pressure levels in the ($p_{\rm gas}$, $T_{\rm gas}$)-profiles where (TiO$_2$)$_{15}$ is the most abundant species are likely to also contain contributions from even larger clusters or the bulk-phase of TiO$_2$.
A general point can be made about the differences of the spectra of different cluster sizes. The absorption cross-sections of larger clusters exhibit a denser forest of stronger lines at shorter wavelengths and more distinct lines at longer wavelengths (Fig. \ref{fig:spec_comparison_size}), when compared to the absorption cross-sections of smaller clusters.  

Additionally there are areas in the ($p_{\rm gas}$, $T_{\rm gas}$) parameter space where not the largest, but the second largest cluster is the most energetically favourable (green area in Fig. \ref{fig:phasediagram_ti_ptprof}, $\sim 1000$K at $10^{-12}$ bar and $\sim 1200-2000$ K at $10^3$ bar). Figure \ref{fig:1d_1e-8bar} shows that also smaller cluster sizes have significant abundances. If this trend continues, not all areas of the parameterspace that are dominated by (TiO$_2$)$_{15}$ ($\sim <1000$K at all pressures) in this work are necessarily dominated by larger clusters if taken into consideration. 
Another caveat of the spectra produced in this work originates from the harmonic oscillator approximation to the vibrational modes of the individual clusters. \cite{Guiu2021HowDust} have shown that anharmonicities 
and non-uniform thermal effects in the vibrational modes have an impact on line position and strength starting already at low temperatures. However, treating all frequency calculations at a full anharmonic level is prohibitively computationally expensive for the scope of this paper. General trends of line intensity or line density increasing with cluster size are not impacted by anharmonicities, enabling qualitative statements on the resulting spectra.\\ 
This work shows, that it is possible to distinguish between the absorption caused by the TiO$_2$ monomer and any (TiO$_2$)$_N$ clusters with $N > 1$, as the absorption cross-sections at wavelengths $\lambda \sim 12 - 25 \mu m$ are up to 3 orders of magnitudes larger for clusters than they are for the monomer. Detection of such an absorption feature with e.g. JWST/MIRI could indicate the presence of TiO$_2$ 
clusters and confirm their role in the formation of cloud condensation nuclei for cloud formation. 
This work shows where in the investigated atmospheres (TiO$_2$)$_N$ 
clusters play a role as absorbing species. 
Other absorbing species could play a larger role or, in the case of the presence of clouds, the atmosphere can become optically thick above the pressure levels where the clusters contribute to the total absorbance. 
Silicate clouds have a feature around 10 $\mu m$ \citep{Wakeford2015AstrophysicsExoplanets}, which is in the vicinity of the shared $\sim 9.5 \mu m$ feature of the clusters from this work. The clusters could therefore explain the existence of a 10 micron feature in the absence of clouds.
We note that, beside (TiO$_2$)$_N$, N=1-15, also other metal oxide clusters exhibit their most intense emissions between 9.5 and 12.5 $\mu$m including clusters of silicates \citep{Zamirri2019WhatGrains} and alumina \citep{Gobrecht2022Bottom-upClusters}. This might represent an additional challenge in identifying and discriminating the cluster species and sizes by IR spectroscopy.
The most interesting ($p_{\rm gas}$, $T_{\rm gas}$) points to probe are the ones contained in the green area in Fig. \ref{fig:phasediagram_ti_ptprof}, where (TiO$_2$)$_{14}$ is the most abundant species, as cluster growth towards larger clusters and the solid phase is inhibited in these regions, but medium sized clusters have already formed and are contributing to absorption with denser and stronger cross-sections. These pressures and temperatures can be probed in the upper atmosphere of a WASP 121b-like exoplanet (black dashed line in Fig. \ref{fig:phasediagram_ti_ptprof}).

\section{Conclusions}
\label{sec:conclusions}
This work has shown that (TiO$_2$)$_N$ 
clusters have unique vibrational absorption spectra. When considering broadening mechanisms of the individual vibrational modes, the lines are broadened beyond JWST/MIRI spectral resolution throughout most of the atmosphere of these exoplanet examples. There are clear differences between the spectra of smaller and larger 
clusters, as increasing size correlates with more lines overall, denser lines at shorter wavelengths and an overall higher cross-section. Different cluster isomers
of the same size have different line positions and intensities, trends in cross-section such as higher cross-sections towards shorter wavelengths. 
Different cluster sizes 
 dominate different areas of the pressure-temperature parameterspace (Fig. \ref{fig:phasediagram_ti_ptprof}),
 including isomers differing from the GM candidate
 dominating the spectra at different points in the studied exoplanet atmospheres. 
 For the WASP 96b-like atmosphere, there is a small difference between the morning and evening terminator spectra due to a higher 
 relative abundance of the (TiO$_2$)$_{14}$ cluster in the latter. For the WASP 121b-like planet, the upper atmosphere is always
 dominated by the TiO$_2$ monomer and the TiO molecule, giving a possible clue as to how deep the atmosphere is probed. When it comes to detecting 
 these 
 building
 blocks that are relevant for cloud formation on their way to become CCN, the atmosphere of an ultra-hot Jupiter such as WASP 121b is better suited for the search when compared to a WASP 96b-like planet.
 
These model spectra of TiO$_2$ clusters can be implemented into radiative transfer models of exoplanet atmospheres to gauge whether their detection with modern instruments is possible. This potential detection or non-detection of the specific fingerprints of the individual clusters could give an indication towards the formation of the
clusters and their role as cloud condensation nuclei in these exoplanet atmospheres. When used with observational data however, the caveats need to be addressed.
The harmonic oscillator approximation used in this work provides a useful and effective means of analysing vibrational spectra. More complex descriptions including anharmonicities, ro-vibrational coupling and non-uniform thermal effects can improve the precision of line positions, but come at much higher computational cost. The changes in line position induced by these effects are small enough (\cite{Guiu2021HowDust} find line shifts of a few percent at wavelengths $\lambda \sim 10\mu m$) to be blurred out at low spectral resolutions, e.g. MIRIs LR mode. Treating these effects can make the data useful at MIRIs medium spectral resolution. The templates this paper provides can be used to model the absorption caused by these CCN-forming clusters at lower spectral resolutions.

\begin{acknowledgements}
J.P.S. acknowledges a St Leonard's Global Doctoral Scholarship from the University of St Andrews and funding from the Austrian Academy of Science. Ch.H. and L.D. acknowledge funding from the European Union H2020-MSCA-ITN-2019 under Grant Agreement no. 860470 (CHAMELEON). D.G. acknowledges funding by the project grant "The Origin and Fate of Dust in Our Universe" from the Knut and Alice Wallenberg Foundation (KAW 2020.0081).
\end{acknowledgements}

% WARNING
%-------------------------------------------------------------------
% Please note that we have included the references to the file aa.dem in
% order to compile it, but we ask you to:
%
% - use BibTeX with the regular commands:
%   \bibliographystyle{aa} % style aa.bst
%   \bibliography{Yourfile} % your references Yourfile.bib
%
% - join the .bib files when you upload your source files
%-------------------------------------------------------------------
\bibliographystyle{aa}
\bibliography{references}

\end{document}